# Perfusion Linearity and Its Applications

Oleg Pianykh, PhD

Abstract
Perfusion analysis computes blood flow parameters (blood volume, blood flow, mean transit time) from the observed flow of contrast agent, passing through the patient's vascular system. Perfusion deconvolution has been widely accepted as the principal numerical tool for perfusion analysis, and is used routinely in clinical applications. This extensive use of perfusion in clinical decision-making makes numerical stability and robustness of perfusion computations vital for accurate diagnostics and patient safety.
The main goal of this paper is to propose a novel approach for validating numerical properties of perfusion algorithms. The approach is based on Perfusion Linearity Property (PLP), which we find in perfusion deconvolution, as well as in many other perfusion techniques. PLP allows one to study perfusion values as weighted averages of the original imaging data. This, in turn, uncovers hidden problems with the existing deconvolution techniques, and may be used to suggest more reliable computational approaches and methodology.
Keywords: Perfusion, Deconvolution, TSVD, Regularization, Computed Tomography.

## 1. Introduction

*Deconvolution* attempts to recover the original convolved function $R(x)$ from the following equation:

$$C(t) = \int_0^t K(t-x) R(x) dx = K \otimes R, \qquad (Eq1)$$

where $C(t)$ is a known (observed) output, and $K(x)$ is a known convolution kernel. In natural phenomena (such as blood flow), all functions in Eq1 are assumed finite and continuous over a finite observation time $t$, $0 \le t \le T < \infty$. If $C(t)$ is observed only at $N$ discrete time intervals $0 = t_0 < t_1 < ... < t_j < ... < t_N = T$, and inter-image time $d_i = t_i - t_{i-1}$ is small enough to assume $R(t)$ and $K(t)$ constant over each $[t_{i-1}, t_i]$, one can numerically approximate Eq1 with a discrete sum in Eq2, based on rectangular integration quadrature[1] [Bronstein]:

$$C_j = C(t_j) = \sum_{i=1}^{j} K(t_j - t_i) R(t_i) d_i, \quad j = 1...N, \quad d_i = t_i - t_{i-1}. \qquad (Eq2)$$

This can be rewritten in matrix format as

$$\mathbf{C} = \mathbf{A_K}\mathbf{R}, \quad \text{matrix } \mathbf{A_k}(i,j) = \begin{cases} K(t_j - t_i)d_i, & 1 \leq i \leq j \leq N \\ 0, & i > j \geq 1 \end{cases},$$

$$\text{vectors} \quad \mathbf{C} = \begin{pmatrix} C(t_1) \\ \ldots \\ C(t_N) \end{pmatrix} = \begin{pmatrix} C_1 \\ \ldots \\ C_N \end{pmatrix}, \quad \mathbf{R} = \begin{pmatrix} R(t_1) \\ \ldots \\ R(t_N) \end{pmatrix} = \begin{pmatrix} R_1 \\ \ldots \\ R_N \end{pmatrix} \quad \text{(Eq3)}$$

leading to a formal solution for the vector $\mathbf{R}$:

$$\mathbf{R} = \mathbf{A_K}^{-1}\mathbf{C} = \mathbf{BC}, \quad \mathbf{B} = \mathbf{A_K}^{-1} \quad \text{(Eq4)}$$

---

[1] It is implied that the error of any discrete approximation depends on the step size $d_i$, choice of sampling points $t_i$, and numerical properties (such as degree of smoothness) of $R(t)$ and $K(t)$. Also, higher order quadratures (trapezoidal, Simpson's) may generally provide more accurate approximations.

In reality, matrix $\mathbf{A_K}$ can be nearly or truly singular, and cannot be inverted directly in Eq4. However, the inverse $\mathbf{B}$ of matrix $\mathbf{A_K}$ can be approximated in numerically-stable manner with various regularization techniques—Truncated Singular Value Decomposition (TSVD) and Tikhonov being the most popular [Tikhonov], [Bronstein], [Hansen]. Nonetheless, regularization of $\mathbf{A_K}$, as complex and non-linear as it might be, does not change the linear nature of Eq4: $\mathbf{R}$ remains a linear function of $\mathbf{C}$. Consequently, any linear function of $\mathbf{R}$ will be a linear function of $\mathbf{C}$ as well.

*Perfusion analysis* quantifies a subject's blood flow through the deconvolution of CT or MR temporal image sequences obtained after a contrast agent injection in the subject's vascular system. The contrast agent (injected into an artery) passes through tissues and organs of interest and changes the observed pixel intensity on the temporal images acquired at $t_i$ time points. As a result, for any given pixel *(x,y)* (due to the physical thickness of CT/MR image slice, representative of a voxel in the 3D tissue volume) one can define a pixel intensity change curve $C_{(x,y)}(t) = C(t)$, as pixel intensity changes at time *t* from the pre-contrast baseline intensity at $t_0 = 0$. It is generally assumed that the blood flow model follows Eq1 [Ostergaard], where function *K(t)* represents the arterial input function AIF[1], *C(t)* – observed contrast change (enhancement) at given pixel *(x,y)*, and *R(t)* – residual function (amount of contrast still present at the *(x,y)* voxel at time *t*); we also assume that the contrast agent is confined to the intravascular space. Then, according to the perfusion deconvolution model (omitting constant scaling coefficients and measurement units), in its discrete form following from Eq2, one computes perfusion blood volume $V_b$ and blood flow $F_b$ as[2]:

$$V_b = \sum_{i=1}^{N} R_i d_i, \quad F_b = R_1 \quad \text{(Eq5)}$$

---

[1] When venous output function (VOF) $C_v(t)$ needs to be taken into account, it is included into *K(t)*.
[2] Inter-image delay time $d_i$ is often the same for all $t_i$, and therefore can be taken outside of summation, as a constant factor $d_i = d$.

For any given voxel, $V_b$ determines the volume of blood, and $F_b$ corresponds to the speed of the blood flow. The third essential perfusion parameter, mean transit time $T_{mtt}$ reflects the average amount of time it takes a particle of contrast agent to pass through the voxel vasculature. According to the central volume principle [Stewart], accepted for perfusion models, $F_b \times T_{mtt} = V_b$, so only two of the three values need to be determined at each pixel, and the third will follow.

However, as one can see from Eq5, both $V_b$ and $F_b$ linearly depend on **R**, and therefore should linearly depend on the original contrast enhancement vector **C**. Indeed, once true or regularized inverse $\mathbf{B} = \mathbf{A_K}^{-1}$ is found in Eq4, one can express

$$V_b = \sum_{i=1}^{N} w_i^V C_i d_i, \qquad F_b = \sum_{i=1}^{N} w_i^F C_i \qquad \text{(Eq6)}$$

where weights

$$w_i^V = w^V(t_i) = \sum_{j=1}^{N} \mathbf{B}(j,i), \qquad w_i^F = w^F(t_i) = \mathbf{B}(1,i) \qquad \text{(Eq7)}$$

In other words, weights $w_i^V$ are column sums of the inverse $\mathbf{B} = \mathbf{A_K}^{-1}$, and weights $w_i^F$ are the first row of **B**.

This leads us to what we call Perfusion Linearity Property (PLP): *Assuming perfusion convolution model ((Eq4), (Eq5)), $F_b$ and $V_b$ are linear combinations of the original values $C_i$.*

Note that PLP follows only from the equations Eq4 and Eq5, and does not assume any particular method of defining or inverting $\mathbf{A_K}$. Therefore, PLP permits one to view *any* perfusion deconvolution algorithm as a *weighted contrast averaging* applied to the original contrast enhancement vector $\mathbf{C} = \{C_1,...C_N\}^T$. The weighting vectors $\mathbf{W^V} = \{w_1^V,...,w_N^V\}^T$ and $\mathbf{W^F} = \{w_1^F,...,w_N^F\}^T$ are location-independent (do not depend on the *(x,y)* voxel location), are derived from the AIF data only, and for each particular perfusion technique are chosen to satisfy the specific algorithm criteria[1]. Therefore, we will define a perfusion-quantifying parameter *P* as *PLP-compliant* if $P = \sum_{i=1}^{N} w_i^P C_i$ and weights $\mathbf{W^P} = \{w_1^P,...,w_N^P\}^T$ do not depend on the voxel position. We will define a perfusion-quantifying parameter *P* as *PLP-norm-compliant* if normalized weighting vector $\mathbf{W^P}/\|\mathbf{W^P}\|$ does not depend on the voxel position. Consequently, we will define a perfusion algorithm as *PLP-(norm-)compliant* if any two of its three principal parameters $\{F_b, V_b, T_{mtt}\}$ are PLP-(norm-)compliant.

Norm-compliant weights $\mathbf{W^P}$ may have norms, depending on voxel coordinates, so norm-compliant definition is more relaxed compared to compliant. However, our analysis will be mainly concerned with relative changes in weighting coordinates $w_i^P$. In this respect norm-compliant and compliant definitions will be identical.
To eliminate any scaling factor irrelevant for our discussion, we will assume all weighting vectors scaled to Euclidean norm: $\|\mathbf{W}\|_2 = 1$ unless stated otherwise.

---

[1] Note an interesting similarity with 3D-rendering techniques, using weighted transfer functions. Perfusion analysis is essentially a transfer function in time.

Although we derived PLP from the deconvolution approach, it can be found in many other popular perfusion algorithms.

## 2. Perfusion Linearity Property in Perfusion Algorithms

A brief review of most popular perfusion analysis methods demonstrates that many of them conform to PLP. Historically, well before perfusion deconvolution was brought into existence, perfusion values were usually computed as [Axel]:

$$V_b = k_V \int_0^T C(t)dt, \quad T_{mtt} = k_T \int_0^T tC(t)dt, \quad F_b = V_b / T_{mtt} \quad \text{(Eq8)}$$

or, in discrete time-sampled format,

$$V_b = k_V \sum_{i=1}^{N} C_i d_i, \quad T_{mtt} = k_T \sum_{i=1}^{N} t_i C_i d_i, \quad F_b = V_b / T_{mtt} \quad \text{(Eq8b)}$$

where $k_v$ and $k_T$ are scale-correcting constants (and $k_T$ does depend on the $V_b$). In this case, $V_b$ is PLP-compliant, and $T_{mtt}$ is PLP-norm-compliant, with weights $w_i^V = d_i$ and $w_i^T = d_i t_i$ respectively (up to scaling factors $k_V$ and $k_T$). This choice of weights had several important advantages:
1. Computational simplicity.
2. Independence on AIF $K(t)$. AIF was used for scaling only (to determine $k_V$ and $k_T$).
3. Uniform weighting for $V_b$, when $d_i$ is constant (the most popular practical choice), and therefore $d_i = d$ can be included into $k_V$. Then an equal weight $w_i^V = 1$ is the only case when the perfusion algorithm does not favor particular time points $t_i$.
4. Independence on time sampling. With $w_i^V = 1$ and $w_i^T = t_i$, changes in image timing would have minimal effect on the perfusion values[1]. This becomes essential in any radiation dose-reduction method when one wants to reduce the number of perfusion images and maintain the same consistent computational approach: Axle's approach is very convenient for this purpose.

---

[1] Assuming $d_i = d$ is still small enough for the discrete summation to be an accurate approximation of the continuous integrals.

After Axel, several variations of slope methods—Patlak perfusion included—were derived from the differential view of the contrast flow [Lee]:

$$\frac{dC(t)}{dt} = F_b \times (C_a(t) - C_v(t)) \quad (Eq9)$$

where the difference between the AIF $C_a(t)$ and VOF $C_v(t)$ is analogous to $K(t)$. Because in the discrete case $dC(t)/dt$ is computed with finite linear differences, equation Eq9 leads to a linear system where $F_b$ and $V_b$ once again are found as linear combinations of $C_i = C(t_i)$, thus conforming to PLP.

The well-known Patlak equation [Lee] [Patlak] [Miles]:

$$\frac{C(t_k)}{C_a(t_k)} = V_r + P_{perm} \frac{\sum_{1}^{N} C_a(t_i) d_i}{C_a(t_k)} \quad (Eq10)$$

where $V_r$ is the relative blood volume and $P_{perm}$ is permeability coefficient also leads to a linear solution, where both $V_r$ and $P_{perm}$ are computed as weighted sums of $C_i$, and the weights depend on $C_a(t)$ only. Thus, Patlak's $V_r$ and $P_{perm}$ are PLP-compliant as well.

Finally, various parametric (curve-fitting) approaches were proposed to find perfusion solutions with certain analytical properties (such as smoothness or exponential decay)[1] [Graz], [Rost]. A smooth curve basis $H = \{H_j(t)\}$, $j = 1,..., N_b < N$, can be fit in the original $C_i = C(t_i)$ sequence with linear regression:

$$C(t) = \sum_{j=1}^{N_b} h_j H_j(t) + e(t) \quad (Eq11)$$

If one considers fit error $e(t)$ as irrelevant noise, then "denoised" $C(t) = \sum_{j=1}^{N_b} h_j H_j(t)$ can be substituted into Axel equation (Eq8) yielding:

$$V_b = k_V \int_0^T \left( \sum_{j=1}^{N_b} h_j H_j(t) \right) dt = \sum_{j=1}^{N_b} h_j \int_0^T k_V H_j(t) dt = \sum_{j=1}^{N_b} h_j H_j^V, \quad H_j^V = k_V \int_0^T H_j(t) dt.$$

$$T_{mtt} = k_T \int_0^T t \left( \sum_{j=1}^{N_b} h_j H_j(t) \right) dt = \sum_{j=1}^{N_b} h_j \int_0^T k_T t H_j(t) dt = \sum_{j=1}^{N_b} h_j H_j^T, \quad H_j^T = k_T \int_0^T t H_j(t) dt.$$

(Eq12)

The constant coefficient vector $\mathbf{h} = \{h_1, ..., h_{Nb}\}^T$ can be found with linear regression from Eq11 as $\mathbf{h} = \mathbf{B_H} \times \mathbf{C}$, where matrix $\mathbf{B_H}$ is derived from the $H_j(t_i)$ values only (does not depend on the contrast $C(t)$). This means that **h** linearly depends on **C**. Because $V_b$ and $T_{mtt}$ in Eq12 linearly depend on **h**, $V_b$ and $T_{mtt}$ linearly depend on **C**. In other words, even when computed through any basis $H = \{H_j(t)\}$, Axel's $V_b$ and $T_{mtt}$ still conform to PLP[1].

---

[1] In essence, this is equivalent to regularization, as one assumes additional analytical properties of the solution to make the solution more stable.

Parametric approach was later revived with deconvolution methods, now applied to $R(t)$ instead of $C(t)$ (see excellent analysis in [Graz] and [Rost]):

$$R(t) = \sum_{j=1}^{N_b} h_j H_j(t),$$

$$C(t_i) = C_i = \sum_{j=1}^{i} C_a(t_i - t_j) R(t_j) = \sum_{j=1}^{N_b} h_j \left[ C_a \otimes H_j(t) \right] = \qquad (Eq13)$$

$$= \sum_{j=1}^{N_b} h_j G_j(t), \quad \text{where} \quad G_j(t) = C_a \otimes H_j(t).$$

But this is an obvious case of Eq11 using $G_j(t)$ instead of $H_j(t)$ and therefore conforming to deconvolution PLP in Eq4 and Eq6. Overall, the only benefit of parametric curve fitting was in proposing yet another way of defining the weighting coefficients $w_i$ in Eq6, approaching the problem from the deconvolution basis angle. However, the entire question of finding the optimal deconvolution basis $\{H_j(t)\}$ has become an art in itself [Graz], significantly contributing to the subjectivity and variability of the perfusion methods, and to the disconnect between the computational and clinical aspects of the analysis.

As a result, all popular perfusion techniques reviewed above are PLP-compliant: $V_b$ and $F_b$ (or $V_b$ and $T_{mtt}$ with Axel-derived methods) are always found as linear combinations of the original contrast values $C_i$. This is expected: PLP holds true for perfusion models because all these models were derived from *linear* flow equations (differential, integral, convolutional, or matrix-based) where equation coefficients were functions of AIF/VOF values *only*. Therefore, for PLP-compliant methods, *the entire question of optimal perfusion algorithm becomes the question of selecting optimal weights $w_i$*. This generalization opens new possibilities for perfusion algorithm analysis and validation.

---

[1] It is easy to show that the linear choice of $H(t)$ leads to "smoothed" formulas for $\mathbf{A_k}$ suggested in [Ostergaard] and [Ostergaard2].

# 3. PLP in Perfusion Deconvolution

## 3.1. PLP in TSVD Deconvolution

PLP can be used to uncover hidden problems in several widely-accepted perfusion algorithms, such as Truncated Singular Value Decomposition (TSVD), which can be formally suggested to inverse singular[1] $\mathbf{A_K}$ in Eq4 by factoring

$$\mathbf{A_K} = \mathbf{U} \times \mathbf{S} \times \mathbf{V} \tag{Eq14}$$

where matrices $\mathbf{U}$ and $\mathbf{V}$ are unitary ($\mathbf{U^T U} = \mathbf{V^T V} = \mathbf{I}$), and matrix $\mathbf{S}$ is diagonal, $\mathbf{S} = \text{diag}(\lambda_1, \lambda_2, ..., \lambda_N)$, $\lambda_1 \geq \lambda_2 \geq ... \geq \lambda_N \geq 0$. The columns in $\mathbf{U}$ and rows in $\mathbf{V}$ are formed by the eigenvectors $u_i$ of $\mathbf{A_K^T A_K}$. Ill-conditioning in $\mathbf{A_K}$ implies that after some threshold index $r < N$, and eigenvalues $\lambda_i$, $i > r$ vanish in absolute value. Therefore, they are considered "noise" and diagonal $\mathbf{S}$ is inverted as $\mathbf{S}_r^{-1} = \text{diag}(1/\lambda_1, 1/\lambda_2, ..., 1/\lambda_r, 0, ..., 0)$. Then matrix

$$\mathbf{B} = \mathbf{V}^T \times \mathbf{S}_r^{-1} \times \mathbf{U}^T \tag{Eq15}$$

in the least-square sense becomes a very close and well-conditioned approximation to $\mathbf{A_K^{-1}}$. To determine the threshold index $r$, 20% of the maximum eigenvalue $\lambda_1$ is widely accepted as a good "generic" cut-off value [Wirestam].

Despite this conceptual clarity, problems with TSVD perfusion deconvolution were empirically observed in many instances, manifesting themselves in poor inter-implementation correlation [Goh], $R(t)$ oscillations [Calamante2], and inconsistent perfusion maps (from our own experience)[Angelos]. Fixing these problems with more elaborate TSVD thresholding (using L-curves [Koh2], block-circulant matrices [Wu2], and regression analysis [Koh]) does not change the essence of the method, but adds computational complexity with no new insights in the original contrast flow process. PLP offers a straightforward and intuitive way of perfusion algorithm validation. From the PLP point of view, TSVD simply proposes yet another approach for computing the weights $\mathbf{W^V}$, $\mathbf{W^F}$ for $V_b$ and $F_b$ ((Eq6), (Eq7)). But because $\mathbf{W^V}$ and $\mathbf{W^F}$ directly relate perfusion measurement to the input $C(t)$ data, they show how each image in a temporal perfusion sequence contributes to the $V_b$ and $F_b$ values.

Consider the graphs on Figure 1, computed from a real CT brain perfusion case ($N = 60$, $T = 60$ sec, $d_i = 1$ sec). We intentionally selected a very clear, motion-free, high-contrast temporal image sequence, resulting in a well-defined gamma-like AIF curve $K(t)$ with no substantial noise in the contrast enhancement phase. We computed $\mathbf{W^V}$ and $\mathbf{W^F}$ for the original sequence of $N = 60$ time points (resulting in $r = 10$ eigenvalue threshold), as well as for the shorter sequence up to the recirculation point ($N = 14$, $r = 3$):

---

[1] Despite the popular belief, $\mathbf{A_K}$ singularity has much more to do with the shape of the AIF function $K(t)$, than with any noise, as we will see later. In fact, noise is likely to make $\mathbf{A_K}$ *less singular*.

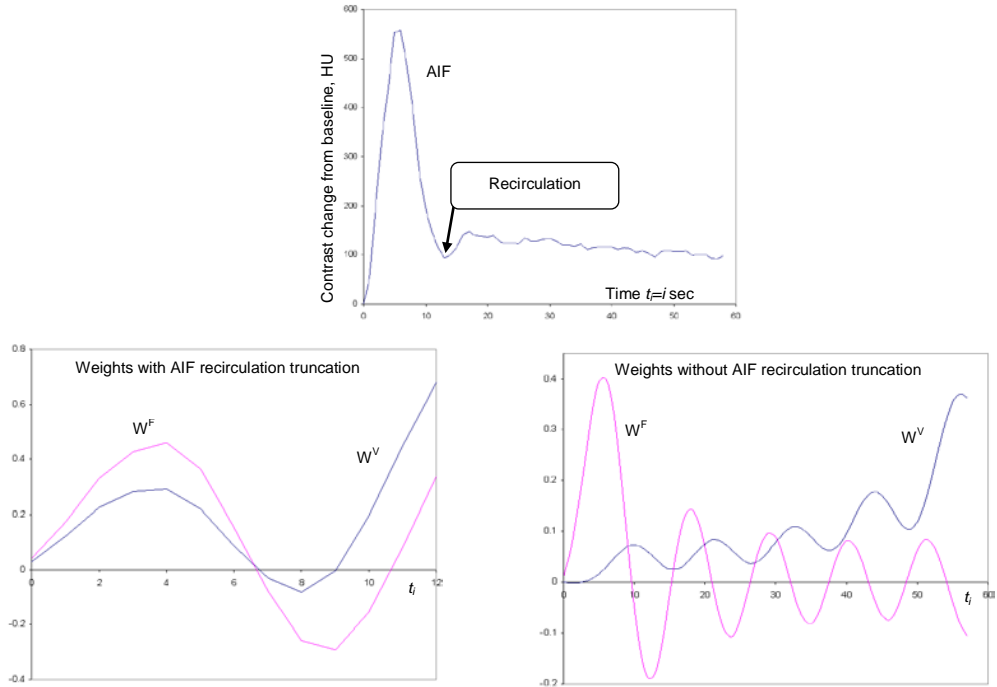

**Figure 1: Original AIF curve $K(t)$ from CT brain perfusion sequence, $t_i = 1...60$ sec, and TSVD-derived weighting sequences $\mathbf{W^V}$ and $\mathbf{W^F}$ - according to equations Eq4 and Eq5. The first set of $\mathbf{W^V}$ and $\mathbf{W^F}$ was obtained from recirculation-truncated AIF. We used standard 20% TSVD eigenvalue cut-off threshold.**

As one can see in Figure 1, TSVD weights $\mathbf{W^V}$ and $\mathbf{W^F}$, found with or without recirculation correction, *do not make any practical sense*:

1. They oscillate severely, making approximately $r$ sign changes, where $r$ is the TSVD eigenvalue truncation threshold.
2. They diverge towards the end. As a result, the most important time points (contrast peak during the first 10 seconds) receive minimal weights (play minimal roles in the $V_b$ and $F_b$ maps), and the least important points towards the end are disproportionally emphasized.
3. They can take negative values—meaning that even the images corresponding to the high-contrast agent intake can have negative contribution to the $V_b$ and $F_b$ values (hard to justify practically, especially for $V_b$).
4. They are completely uncorrelated with the AIF shape.
5. They are severely affected by recirculation truncation, or by any truncation in general (choice of $N$ time points or choice of total scan time $T$): $\mathbf{W^V}$ and $\mathbf{W^F}$ for the full AIF ($N = 60$) have nothing in common with $\mathbf{W^V}$ and $\mathbf{W^F}$ for the recirculation-truncated AIF ($N = 14$). This makes the choice of $N$ and $T$—often performed manually and another source of numerical instability—capable of completely changing the TSVD perfusion outcomes.

These observations mean that TSVD-based perfusion analysis has serious flaws, which cannot be fixed by more intricate approaches to the eigenvalue thresholding. Moreover, as our numerical experiments indicate, oscillating and diverging patterns in PLP weights are very common for TSVD deconvolution. In other words, TSVD perfusion solutions *can be very unstable and divergent even without substantial noise or artifacts in the original data.*

What causes these problems? TSVD itself.

TSVD eigenvectors $u_k$ are known to oscillate as $k$ increases, approximately with $k$ sign changes in $u_k$ [Hansen]. These oscillations inevitably propagate into the **B** matrix (Eq15), and then into the PLP weights $\mathbf{W^V}$ and $\mathbf{W^F}$. One can take an "ideal" noise-artifact-recirculation-free AIF curve, such as $K(t) = \gamma(t) = t^3 e^{-t/1.5}$ (often used in numerical perfusion simulations [Ostergaard]) and observe the same oscillation phenomenon, as shown on Figure 2.

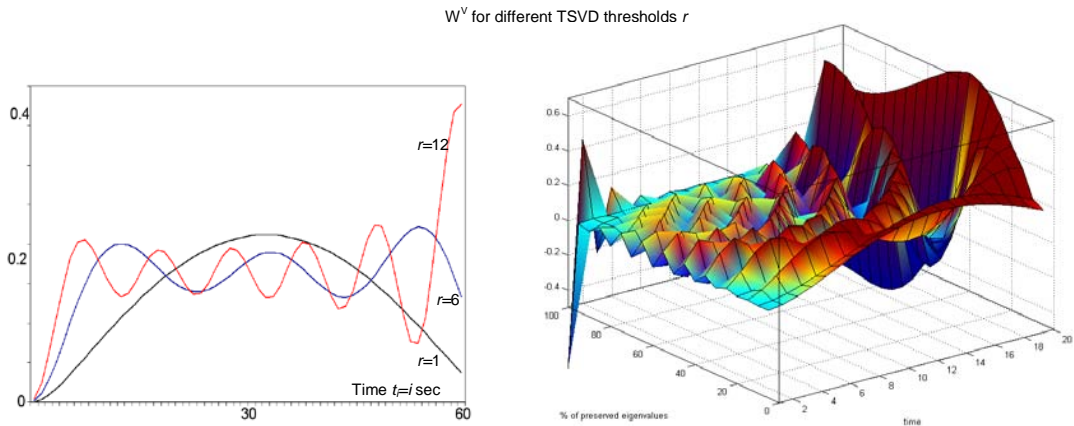

**Figure 2**: *Left*: $V_b$ weights $\mathbf{W^V}$ computed after keeping first $r = 1$, $r = 6$, and $r = 12$ eigenvalues $\lambda_i$ in TSVD deconvolution based on AIF $K(t) = t^3 e^{-t/1.5}$ (sampled at N = 60 time points $t_i = 1...60$). Note that the number of sign changes in each $\mathbf{W^V}$ is roughly equal to the number $r$ of preserved eigenvalues. Note also the diverging of weights for large $t_i$ when more eigenvalues are preserved.

*Right*: More complete image of $\mathbf{W^V}$ as a function of preserved eigenvalues and time (now sampled at N = 20 time points $t_i = 1...20$). The case of 100% preserved eigenvalues corresponds to all $r = 20$ all eigenvalues kept, and no regularization applied. Note the high oscillation even for low eigenvalue thresholds $r$ (high regularization), and diverging pattern for late time points $t_i$. Using longer sampling sequence of $t_i = 1...60$ would only worsen this pattern.

One can see that the local extrema in TSVD-derived $\mathbf{W^V}$ change their count and locations depending on the cutoff threshold $r$, randomly favoring different images in the original $C(t_i)$ sequence. The same problem can be shown for $F_b$ weights $\mathbf{W^F}$. As a result, the quality of TSVD perfusion analysis is severely affected by the number $r$ of preserved SVD eigenvalues (eigenvectors) and depends on $r$ more than on anything else. Theoretically, one wants to preserve as much data (eigenvalues) as possible, but practically, keeping more eigenvalues with TSVD means introducing more oscillations in the $\mathbf{W^V}$ and $\mathbf{W^F}$ weighting sequences, making the values of $V_b$, $F_b$, and $T_{mtt} = V_b/F_b$ more and more meaningless. Moreover, increasing the total number of perfusion time points $N$ plays the same role: it means increasing the total number of eigenvalues, and likely increasing the number of preserved eigenvalues, which can lead to very unstable solutions. Ironically, to make TSVD numerically-stable, one needs to keep the size of perfusion series $N$ as low as possible, which generally leads to less accurate measurements.

Finally, weighting sequences obtained for different eigenvalue thresholds $r$ appear to be completely uncorrelated with each other (Figure 2), which means that different thresholding strategies in TSVD perfusion algorithms will lead to completely different perfusion values of $V_b$, $F_b$ and $T_{mtt}$. This explains the well-known disaccord between perfusion values measured in different commercial software: although they claim to use the same TSVD algorithm, their outcomes depend on the eigenvalue thresholding more than on the original data.

### *3.2.  PLP in Tikhonov Deconvolution*

As we already mentioned, TSVD is not the only regularization technique that can be applied to solving Eq4. Let's use PLP to consider another popular (and less computationally-expensive) method—Tikhonov regularization [Tikhonov], [Koh2], [Calamante2]—where the original matrix $\mathbf{A_K}$ is conditioned with linear constraint $\mathbf{L}$ to compute $\mathbf{B}$ as:

$$\mathbf{B} = (\mathbf{A_K^T A_K} + \alpha \mathbf{L^T L})^{-1} \mathbf{A_K^T} \qquad (Eq16)$$

The most common choices for the matrix $\mathbf{L}$ are either identity matrix $\mathbf{I}$ or first-order derivative matrix[1] $\mathbf{D}$—although others have been suggested [Koh2]. Regularization parameter $\alpha \geq 0$ plays the same role as TSVD truncation threshold $r$: higher $\alpha$ corresponds to more regularized solutions (similarly to lower $r$, or higher $\lambda_r$). The general theory behind the optimal $\alpha$ selection (L-curves in particular) can be applied, but as we already mentioned, it optimizes $\alpha$ with respect to the residual norms and not the expected properties of the contrast flow.

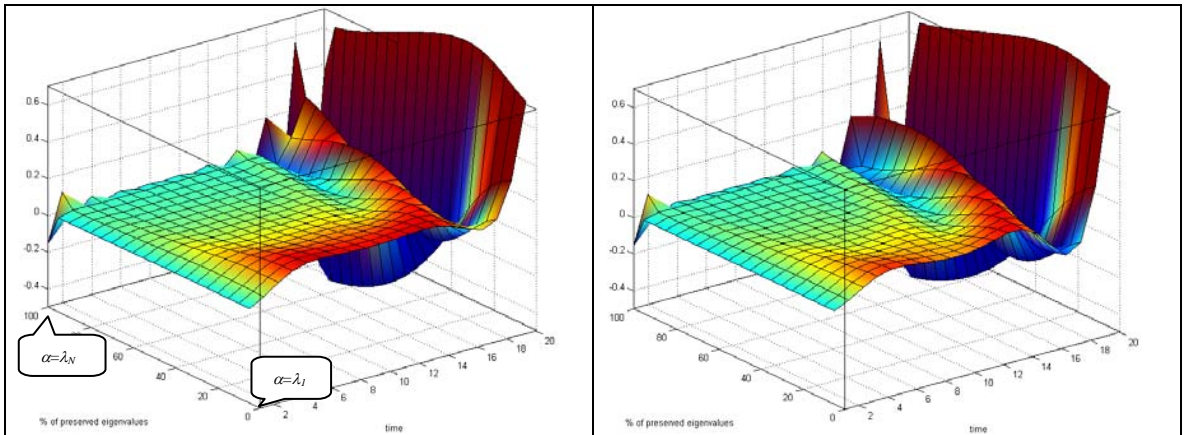

**Figure 3:** Comparing $V_b$ weights $\mathbf{W^V}$ computed with Tikhonov regularization for AIF $K(t) = t^3 e^{-t/1.5}$ (sampled at N = 20 time points $t_i$ = 1...20), using $\mathbf{L} = \mathbf{I}$ (left) and $\mathbf{L} = \mathbf{D}$ (right). Parameter $\alpha$ was set to

---

[1] $D(i,i) = 1$, $D(i, i + 1) = -1$, 0 otherwise.

**the SVD eigenvalues $\lambda_r$, $r = 1...N$. Note that despite the change in the regularization matrix L, both methods behave very similarly—including severe divergence at the end (high $t_i$).**

The graphs on Figure 3 show $V_b$ weighting coefficients $\mathbf{W^V}$ found for different degrees of regularization $\alpha = \lambda_r$ ($r = 1…20$). Compare it to the Figure 2, right. As we already know, TSVD exhibits extremely unstable behavior (with respect to the regularization threshold). On the other hand, $V_b$ weights $\mathbf{W^V}$ obtained from Tikhonov regularization are more oscillation-free [Calamante2]—yet they have problems of their own. Divergence in Tikhonov's $\mathbf{W^V}$ towards the higher $t_i$ exceeds that of TSVD, manifesting itself even for very regularized (high $\alpha$) solutions. At the same time, the main part of $\mathbf{W^V}(t_i)$ remains nearly flat (with an absolute value close to 0), which means that the bulk of the $C(t)$ data—contrast peak included—is essentially ignored. The change in the regularization matrix $\mathbf{L}$ seems to make no improvement. As a result, $V_b$ values obtained from such less-oscillating Tokhonov $\mathbf{W^V}$ make little sense, reflecting more of regularization method artifacts than the true blood volume. The same can be shown for the flow $F_b$ (weights $\mathbf{W^F}$) and therefore will follow for $T_{mtt}$.

To conclude, as we have discovered with Eq6, what really matters in perfusion deconvolution are the PLP weights: only the weight values are needed to compute blood volume $V_b$ and flow $F_b$ from the original data $C(t)$. This makes PLP analysis instrumental in perfusion algorithm selection and validation. Regardless of deconvolution regularization techniques, we expect the weighting coefficients $\mathbf{W}$ to be smooth and non-oscillating, corresponding to a generally smooth and monotone contrast flow. *Any oscillations, high slopes, or negative values in weighting coefficients are hard to justify clinically and pragmatically*. But TSVD and Tikhonov regularization methods fail to satisfy these criteria. As a result, *smooth weights $w_i$ are obtained from the smooth original AIF data $K(t)$ via numerically-unstable deconvolution algorithms*.

### 3.3. $A_K$ Singularity and AIF Shape

The main justification behind all perfusion regularization approaches was dealing with $\mathbf{A_K}$ singularities (Eq4). These singularities were commonly attributed to measurement noise; and regularization techniques were meant to help with denoising.
However, $\mathbf{A_K}$ singularity and deconvolution instability may have nothing to do with noise or artifacts (see Figure 2). In fact, even smooth gamma-shaped functions $K(t)$ can produce ill-conditioned $\mathbf{A_K}$ in Eq3.
The relationship between $K(t)$ and singularity of $\mathbf{A_K}$ can be demonstrated in many ways. First, if $K(0)$ vanishes to 0 (which is often the case), then so do the diagonal elements in $\mathbf{A_K}$ (see Eq3), already making $\mathbf{A_K}$ singular[1]. More intricate observations can be derived from Figure 4, demonstrating how the shape of an ideal, noise-free AIF $K(t)$ can affect the stability of its perfusion deconvolution.

---

[1] Ironically, this is the case when noise in $K(t)$ can prevent $\mathbf{A_K}$ from having 0 on the diagonal, making it *less* singular!

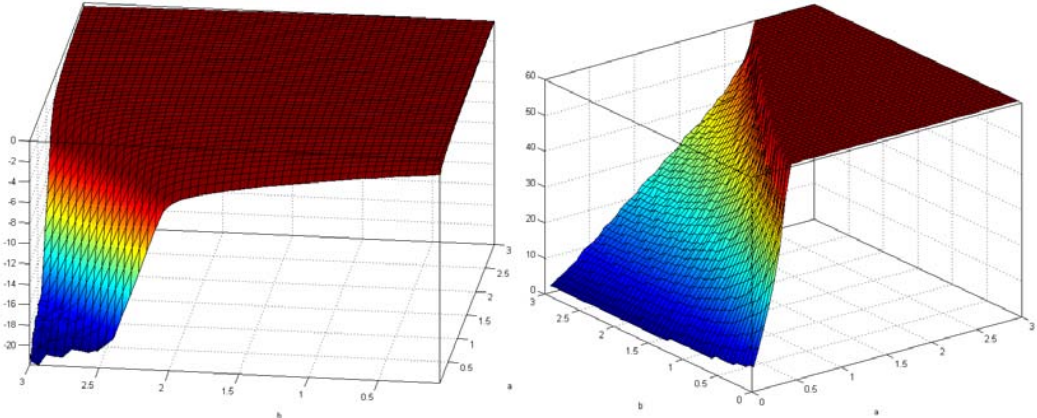

**Figure 4:** *Left*: Plot for $log_{10}(\lambda_N/\lambda_1) = log_{10}(\lambda_{min}/\lambda_{max})$ for SVD eigenvalues of matrix $\mathbf{A_K}$. Matrix $\mathbf{A_K}$ corresponds to $K_{a,b}(t) = t^a e^{-bt}$ AIF, sampled for $t_i = 1...60$ ($N = 60$), for different values of *a* and *b*.

*Right*: Plot for the 20% cutoff index $r(a,b)$, such that $\lambda_r >= 0.2 \times \lambda_{max} > \lambda_{r+1}$.

As one can see from Figure 4 (left), even the shape of $K(t)$ can substantially affect numerical stability of perfusion deconvolution. For example, large *b* can considerably limit the numerical rank of $\mathbf{A_K}$, creating $\lambda_N$ closer to 0. Large values of *b* in $K_{a,b}(t) = t^a e^{-bt}$ correspond to fast-decaying AIF curves (or long scan times *T*), which lead to faster-decaying SVD eigenvalues. However, fast decay in eigenvalues means lower cutoff index *r* (if *r* corresponds to $0.2\lambda_1$ for instance), fewer preserved eigenvalues, and fewer oscillations in weights **W**. In other words, "good" non-oscillating **W** are likely to correspond to intrinsically ill-conditioned $\mathbf{A_K}$ when ill-conditioning is rooted in the shape of $K(t)$ rather than in measurement noise or artifacts.

This conclusion is also supported by Figure 4 (right), showing how the "20% cutoff" index *r* changes with different choices of the AIF shaping parameters *a* and *b*. The surface on this plot consists of two major areas: a large flat "plateau" (where 20% cutoff does not eliminate any of the $N = 60$ SVD eigenvalues) and a rather steep "wall" corresponding to eigenvalue truncation.

For the "plateau" area, all eigenvalues are preserved, which means that matrix $\mathbf{A_K}$ is well-conditioned to be inversed "as is." In other words, TSVD is not needed and is not effectively applied.

For the "wall" area, TSVD is used to keep only the first $r < N$ eigenvalues, but please note how steep the wall is—especially for small values of *b* (slow decay in AIF). Essentially, very subtle changes in the AIF shape variables *(a,b)* can lead to a substantial change in the TSVD truncation threshold *r*: changing the number of oscillating eigenvectors and seriously affecting the outcomes of the perfusion deconvolution. Note that for any constant *c*, $K_{a,b}(ct) \sim K_{a,cb}(t)$ and changing the total scan time *T* (set to $N = 60$ seconds in our case) will simply mean using another value of *b* on the same plot. Thus, Figure 4 (right) can be used to judge the stability of any TSVD deconvolution where the AIF curve can be closely approximated by gamma-like $K_{a,b}(t)=t^a e^{-bt}$ regardless of the total scan time.

To conclude, matrix-regularization approach to solving Eq2 has problems of its own, creating numerically-unstable results even for smooth, noise- and artifact-free input functions. Moreover, the entire concept of regularization contradicts the original convolution model of perfusion flow in Eq3. Matrix $\mathbf{A_K}$, regularized with TSVD or Tikhonov methods, loses its original Toeplitz convolution-specific form in Eq3—that is, the regularized matrix generally does not correspond to any flow deconvolution equation or kernel *K(t)*. When we diverge from the original *K(t)*, we diverge from the physical convolving nature of perfusion flow, and we inevitably arrive at diverging—and therefore meaningless—weighting sequences.

One may wonder why deconvolution, so long popular in perfusion applications, was producing reasonable perfusion maps. In our opinion, this became possible mainly because the original *C(t)* data had enough contrast to tolerate suboptimal and even erroneous choice of **W**. As many radiologists know, good perfusion maps were always attributed to good perfusion algorithms, and bad perfusion maps were always blamed on "noise," "low contrast," and "artifacts." High doses of contrast and radiation (for CT scans) cushion these maps from diverging and oscillating weights. But high contrast and exposure in perfusion images may be very costly for the patients, and should never be used to compensate for the inherent perfusion algorithm deficiencies.
Deconvolution regularization deficiencies can be observed in many current perfusion software packages where AIF selection can be done in the most anatomically-incorrect locations (on bones for instance) and still result in visibly sound perfusion maps. A cleaner approach to perfusion visualization is imperative.

## 4. Enhancing Perfusion Analysis with PLP

### 4.1. Principal Component Perfusion

The PLP view of perfusion computations can lead us to the new approaches for perfusion measurements.
Fundamentally, perfusion analysis is meant to qualify the passing of the contrast agent through the tissues of interest. Consequently, optimal perfusion analysis method has to do this in the most visible and numerically-stable way. But according to PLP, the only constant assumption we seem to make is that perfusion values are weighted sums of the original contrast enhancement values $C_i$. Therefore, for optimized perfusion analysis, one needs to select the weights **W** from Eq6 in such a way that the perfusion map would show as much contrast variance as possible. This means that *the vector* **W** *should be chosen as the first principal component of the perfusion image sequence.* Therefore, we define the First Principal Component (FPC) parameter $P_{FPC}$ as:

$$P_{FPC} = \sum_{i=1}^{N} w_i^{FPC} C_i \qquad (Eq17)$$

where weights $w_i^{FPC}$ are the coordinates of the first principal component. This choice of $\mathbf{W^{FPC}} = \{ w_1^{FPC}, \ldots, w_N^{FPC} \}^T$ guarantees the optimal (in least-square sense) representation of contrast-related variance in a single parameter $P_{FPC}$.
One can arrive at Principal Component Analysis (PCA) in Eq17 from a few different angles as well. Consider Eq11 and Eq13. They were derived from different assumptions (Axel model in Eq11 and convolution in Eq13), but they all agree in representing *C(t)* as a linear combination of $N_b$ basis functions where $N_b$ is expected to be substantially lower than the total perfusion image count $N$. The least-square optimal choice of the basis is given by PCA. In many ways, PCA (similar to PLP) directly follows from the original linearity of perfusion modeling, regardless of the underlying model assumptions.

Unlike "volume," "flow," and "MTT" the value of $P_{FPC}$ is not related to any pharmokinetic model. However, it only seems to be an advantage. First of all, $P_{FPC}$ depends only on the perfusion data with no models attached, which makes it universal and independent from often subjective model assumptions. This, for example, explains increasing applications of PCA in perfusion CAD (such as 3TP method [3TP] and its derivatives). Second, with a variety of current perfusion algorithms (poorly correlated with each other, as we have demonstrated above), "volume," "flow," or "MTT" interpretations have become highly inconsistent and model-specific. In fact, they are nothing but the names for specific model parameters, and "blood volume" from one model may have nothing to do with the "blood volume" from another. We should be much more concerned with our ability to optimally see the effects of the contrast perfusion, and this is exactly what the value of $P_{FPC}$ provides.

The other advantages of using $P_{FPC}$ map follow from the PCA properties:
1. One does not have to define AIF/VOF points. Not only does this eliminate the subjectivity of AIF/VOF point definition, it also totally automates computation of $P_{FPC}$.
2. Consequently, FPC analysis can be done specifically for a selected region of interest and not for the entire image. Because the analysis does not depend on the AIF/VOF points, we can limit our maps to the areas we are interested in—such as tumors. AIF/VOF points do not have to be in the field of view; moreover, FPC can process cases where AIF/VOF points were not originally scanned or are generally missing in the image data. Local/regional FPC analysis can also be beneficial when one wants to exclude certain structures—for instance, large blood vessels.
3. PCA is extremely insensitive to noise and is often used as a robust noise-removal technique. Because $P_{FPC}$ will underline as many changes in contrast flow as computationally possible, one can use smaller amounts of contrast and less radiation exposure.
4. For the same reason, FPC can be done without any initial image smoothing or noise filtering. Combined with PLP's "weighted" view of perfusion analysis, it produces a really fast computational method.
5. PCA has already been applied to certain problems of perfusion imaging such as minimization of recirculation artifacts [Wu]. Now it can be done in a single method.

If one compares FPC weights to those from perfusion deconvolution, it is easy to see that FPC weights follow a much more stable pattern. This can be seen with our previous example of CT brain perfusion from Figure 5 (we used only $\mathbf{W^V}$, but $\mathbf{W^F}$ will show similar patterns):

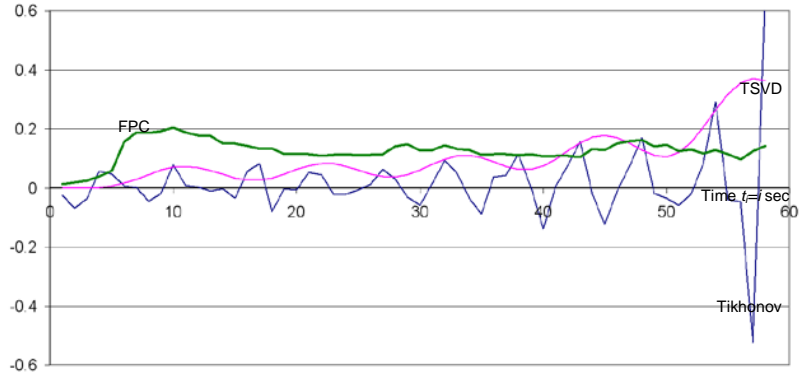

**Figure 5: Weighting series for CT brain perfusion case: PCA-derived FPC series $W^{FPC}$, TSVD-derived $W^V$, and Tikhonov-derived $W^V$. Note that while the FPC series favors the first temporal images (when most of contrast agent is passing through the tissue), TSVD and Tikhonov $W^V$ oscillate inconsistently and diverge at the end, favoring the last and the least important images.**

Consequently, $W^{FPC}$ coefficients produce very clear $P_{FPC}$ image (Figure 6):

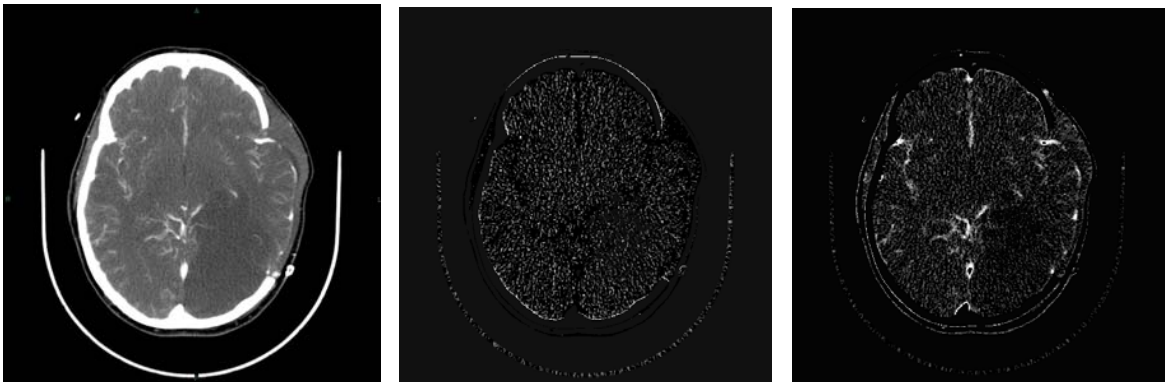

**Figure 6: Original image, TSVD $V_b$ map and PCA first principal component $P_{FPC}$ map. Note the missing vascular structure on the TSVD map: it received low weights and was essentially suppressed. This is one of the cases when TSVD problems become very visible.**

As a result, the FPC map provides a much better qualitative view of perfusion compared to the less-stable deconvolution. Therefore, in our opinion, it is essential to include FPC maps in any visual perfusion analysis. Considering the principal component above first may be beneficial as well, but we will leave it for a separate discussion.

### *4.2. Perfusion Value Orthogonality and Consistency*

The use of weighting sequences **W** for all PLP-compliant perfusion parameters brings up the subject of perfusion parameter independence. How many perfusion parameters (maps) does one need to access perfusion phenomena?

The answer for PLP-conforming parameters (algorithms) is straightforward: *perfusion parameters are linearly independent only if their PLP weights are orthogonal. Perfusion parameters with highly-correlated weights are redundant*. Undoubtedly, parameter orthogonality should be present in any sound PLP-compliant algorithm.

Axel's choice of $w_i^V = 1$ and $w_i^T = t_i$ (in Eq8b, assuming equidistant timing) gives a good example of perfusion *orthogonality*: regardless of sampling, correlation between the discreet weighting vectors $\mathbf{W^V}$ and $\mathbf{W^T}$ is zero, which means that Axel's $V_b$ and $T_{mtt}$ maps will indeed provide different (linearly independent, orthogonal) views of the perfusion phenomena. Using $\mathbf{W}$ from PCA components will produce orthogonal parameters as well by PCA definition. TSVD and Tikhonov's weighting sequences, as our earlier studies indicate, are clearly not orthogonal.

### *4.3. Continuous View of PLP Weighting and Parameter Consistency*

It's time to recall that contrast flow is a continuous phenomenon; and even though all our measurements remain discrete, we should keep the continuity in mind. The definition of PLP weighting naturally extends to the continuous case: we can define the perfusion parameter *P* as *PLP-compliant in a continuous sense* if there exists a continuous weighting function *W(t)* defined on [0,∞), $\int_0^\infty W^2(t)dt < \infty$ , such that for any pixel location *(x,y)*,

$$P = \int_0^T W(t)C(t)dt \qquad \text{(Eq17)}$$

In other words, *W(t)* (identically to weighting vector $\mathbf{W}$ in the discrete case) defines how a contrast change *C(t)* at each time *t* contributes to the value of *P*.

Does *W(t)* always exist? The answer entirely depends on the perfusion analysis method. For example, for Axel's approach (Eq8), it exists and defines as $W_b(t) = 1$ for $V_b$, and $W_{mtt}(t) = t$ for $T_{mtt}$. Moreover, the existence of *W(t)* is an extremely important prerequisite for any sound perfusion analysis method. In essence, it means that all discrete PLP weighting vectors $\mathbf{W}_\sigma$ obtained for different choices of time sampling sequences σ = {$t_0, t_1, ..., t_N$} are nothing more than sampled *W(t)* to which they converge as the sampling step $d = \max_i |t_i - t_{i-1}|$ vanishes to 0. Consequently, this means that discrete computation of *P* will be stable and error-tolerant with respect to the choice of time sampling sequence σ, because with sufficiently small choices of *d* discrete estimates of *P*, regardless of σ, can be made arbitrarily close to the "true" continuous value of *P* in Eq17. Therefore, for any given perfusion parameter *P*, we would like to define the uniform convergence of discrete $\mathbf{W}$ to a continuous *W(t)* as a parameter *consistency* property. By definition, only consistent parameters can tolerate changes in time sampling sequences. This is extremely important for practical applications. With any general theory on perfusion time sampling lacking, different institutions apply different time-sampling strategies. Also, only consistent parameters can be used for perfusion experiment design, as we will soon see.

From our previous analysis, we can conclude that Axel's parameters are consistent. The PCA parameter consistency can be also demonstrated under some reasonable conditions on *C(t)*. Just think about *d* being small enough so that for any *t* in ($t_{i-1},t_i$) its *C(t)* can be accurately linearly-interpolated from *C($t_{i-1}$)* and *C($t_i$)*; then, principal components of the image set won't change at all. However, as we have seen, TSVD and Tikhonov's approaches to perfusion analysis produce *inconsistent* parameters.

With the number of perfusion parameters increasing over the past few years, their orthogonality and consistency should become one of the main requirements for their use in clinical applications.

### *4.4. Optimal Scheduling and Dose Reduction*

The science of optimal experiment planning attempts to build an ideal experiment design that will produce the most reliable measurements [Fedorov]. In the case of perfusion, we are interested in selecting the optimal temporal sequence $t_i$, $0 = t_0 < t_1 < ... < t_j < ... < t_N = T$ so that:
1. *N* is minimal. This means taking as few images as possible to shorten the total scan time and to reduce patient radiation exposure in CT perfusion.
2. $t_i$ are distributed in the most optimal way, which guarantees the most accurate measurement (perfusion values).

However, even simple dose-reduction methods such as uniform image subsampling (increasing sampling time interval $d_i = t_i - t_{i-1}$ to reduce total image count *N*) can be greatly affected by the choice of the perfusion algorithm [Rost]. This is directly related to our definition of parameter consistency. Algorithms with low consistency (such as TSVD) tend to produce high random peaks in their weighting sequences **W**. Therefore, if we use them with optimal experiment design—to reduce the number of temporal images—the outcome will entirely depend on how the image-removal pattern interacts with the random weight oscillations. In short, weight-*inconsistent* algorithms make optimal perfusion scheduling impossible (image-reduction strategies included). Figure 7 and Figure 8 illustrate this for Tikhonov and TSVD deconvolution.

For each algorithm, we used our initial 60-timepoint CT perfusion sequence to compute blood volume weights $\mathbf{W^V}$ (the results for $\mathbf{W^F}$ look similar). Then we analyzed three image-reduction strategies; uniformly removing three of every four images, truncating the whole image sequence to the first 14 images (up to the AIF recirculation point), and replacing the images with the highest contrast content ($t_i = 5$ and $t_i = 6$ sec, corresponding to AIF peak) by their linear interpolations from the neighboring images ($t_i = 4$ and $t_i = 7$ sec).

This time, Tikhonov deconvolution regularization ($\alpha$ was set to the 20% TSVD cutoff eigenvalue, L = I) produced the most incoherent $\mathbf{W^V}$, as can be seen on Figure 7. The weighting sequences came out to be completely uncorrelated and diverging towards the end.

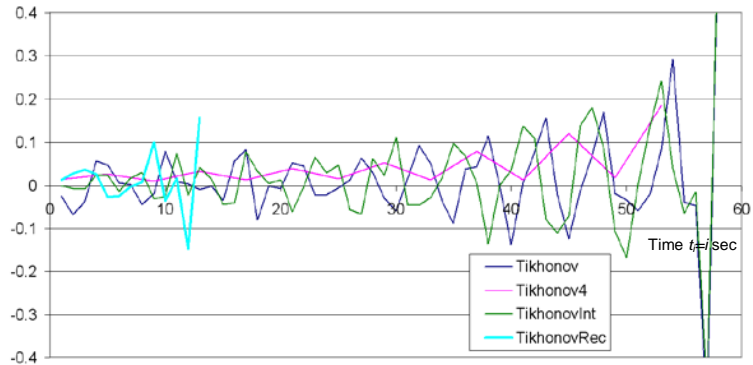

**Figure 7: Experiment planning with Tikhonov deconvolution ($\alpha$ is set to the 20% cutoff eigenvalue $L = I$). Series "Tikhonov" is the $\mathbf{W}^V$ series for a brain CT sequence; "Tikhonov4" was obtained after keeping only each 4-th image, "TikhonovInt"—after replacing AIF peak images 4 and 5 by their linear interpolation, and "TikhonovRec"—after considering only the first 14 images for recirculation correction.**

TSVD deconvolution regularization (same 20% eigenvalue cutoff) produced smoother $\mathbf{W}^V$ (Figure 8) yet oscillating, diverging, and completely uncorrelated to be used for any image-reduction strategies. The only close match was observed between the original (TSVD) and interpolated (TSVDInt) sequences, but in fact this is the case when we would like to see some difference in the weights. The interpolation changed the most contrast-reach images, and we would expect $\mathbf{W}^V$ to reflect this change at the time points where it occurred ($t_i = 5$ and $t_i = 6$ sec).

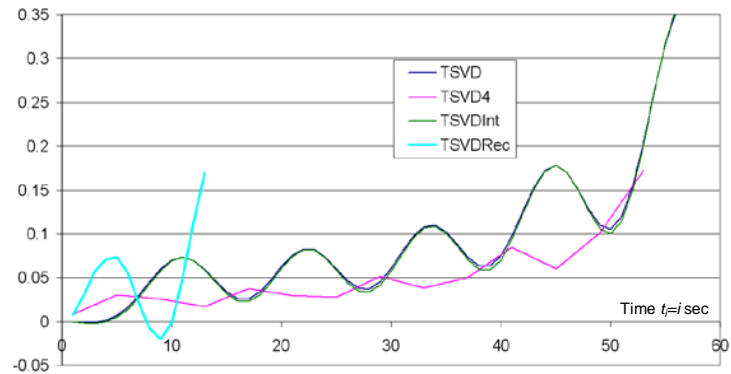

**Figure 8: Experiment planning with TSVD deconvolution (20% cutoff eigenvalue). Series "TSVD" is the $\mathbf{W}^V$ series for a brain CT sequence; "TSVD4" was obtained after keeping only each 4-th image, "TSVDInt"—after replacing AIF peak images 4 and 5 by their linear interpolation, and "TSVDRec"—after considering the first 14 images only, for recirculation correction.**

PCA-derived weights $\mathbf{W}^{FPC}$, contrary to TSVD and Tikhonov deconvolution, demonstrated consistency and proper response to the scheduling changes. First of all, they remained very similar after removing three of every four images ("PCA4") and after recirculation correction of AIF ("PCARec"). As for the AIF contrast peak interpolation ("PCAInt"), it affected $\mathbf{W}^{FPC}$ exactly where it happened ($t_i = 5$ and $t_i = 6$ sec) and nowhere else. Thus, *PCA weight distribution $\mathbf{W}^{FPC}$ is consistent and can be used for image-reduction strategies*.

Also, note the similarity between the $\mathbf{W^{FPC}}$ data and the original AIF shape (Figure 1, top). High contrast images (around AIF peak) naturally correspond to large contrast variance and higher $\mathbf{W^{FPC}}$ values.

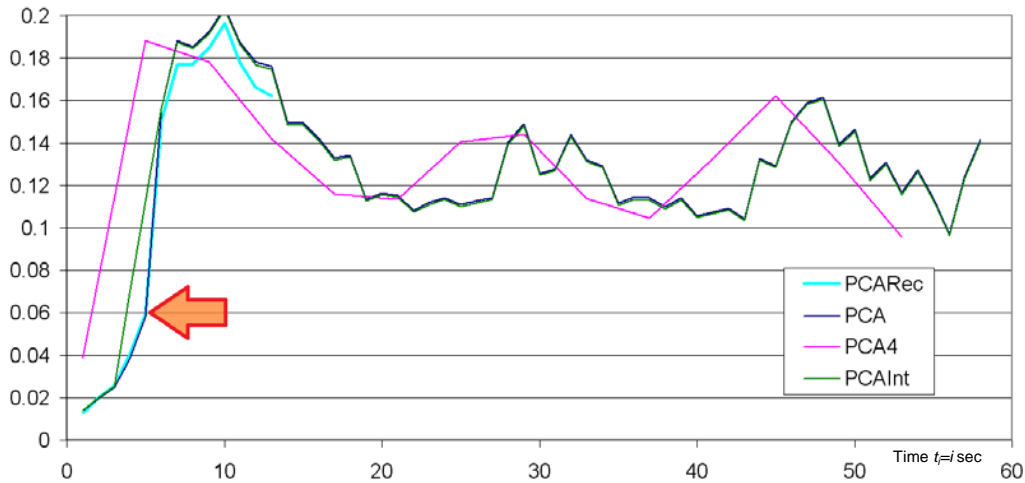

**Figure 9: Experiment planning with FPC map. Series "PCA" is the $\mathbf{W^{FPC}}$ series for a brain CT sequence; "PCA4" was obtained after keeping only each 4-th image, "PCAInt" – after replacing AIF peak images 4 and 5 by their linear interpolation, and "PCARec" – after considering the first 14 images only, for recirculation correction.**

As one can conclude, PCA weights $\mathbf{W^{FPC}}$, used to build the proposed optimal FPC map, behave in much more stable and meaningful way compared to regularization-based deconvolution weighting.

### 4.5. Future Research: PCA Eigenvalues

Eigenvalues $\Lambda = \{\lambda_1, \lambda_2, ..., \lambda_N\}$, produced by PCA, lead to another interesting direction in perfusion analysis. Consider the following reasoning. In a normal region of interest, contrast enhancement will be mainly produced by the large vessels (such as arteries) roughly corresponding to the same time $T_0$ after the contrast injection. Peaking around a single time point makes $C(t) = \{C_1,..., C_N\}^T$ vectors essentially one-dimensional—one coordinate at $t = T_0$ dominates the others. This means that $\lambda_1$ will have a large relative value

$$E_\Lambda = \frac{\lambda_1}{\sum_{i=1}^{N} \lambda_i} \qquad (Eq18)$$

which in our experiments was as high as 0.5 for normal brain images. On the other hand, in the presence of an abnormally-perfusing organ (such as a tumor), $C(t)$ should exhibit another noticeable contrast enhancement at some different, more delayed time $T_1$. This will increase the intrinsic dimensionality of the $C(t)$ vector space, reducing the relative value of $\lambda_1$ and possibly contributing to a higher value of another $\lambda_K$. In PCA terms, $\lambda_i$ reflects true data dimensionality, and abnormal tissues may add other dimensions compared to the normal. As a result, the distribution of eigenvalues $\{\lambda_1, \lambda_2, ..., \lambda_N\}$ can be used to analyze the uniformity and normality of perfusion flow, extending what we used to expect from $T_{mtt}$. We leave this topic for further investigation.

# 5. Conclusion

In this paper, we introduced Perfusion Linearity Property (PLP), which naturally follows from the linearity of many well-known perfusion flow models. For PLP-compliant perfusion parameters (such as $V_b$ and $F_b$ in deconvolution methods), perfusion parameter selection is nothing but the selection of PLP weights $w_i$.

This generalization permits to analyze, validate, and compare numerical properties of different perfusion methods via their PLP weighting sequences. This also introduces the concepts of perfusion parameter orthogonality and consistency, which should be used for optimal perfusion algorithm selection.

Finally, PLP can be used to develop new perfusion visualization approaches (such as PCA maps), study dose reduction problems, and can potentially lead to more insightful perfusion quantification.